# Large negative magnetoresistance and pseudogap phase in superconducting A15-type La$_4$H$_{23}$

Jianning Guo [1, †], Dmitrii Semenok[2, †, *], Grigoriy Shutov [3, †], Di Zhou[2], Su Chen[1], Yulong Wang[1], Toni Helm[4], Sven Luther[4], Xiaoli Huang[1, *], and Tian Cui [1,5]

[1] *State Key Laboratory of Superhard Materials, College of Physics, Jilin University, Changchun 130012, China*
[2] *Center for High Pressure Science and Technology Advanced Research (HPSTAR), Beijing*
[3] *Skolkovo Institute of Science and Technology, Skolkovo Innovation Center, 3 Nobel Street, Moscow 143026, Russia*
[4] *Hochfeld-Magnetlabor Dresden (HLD-EMFL) and Würzburg-Dresden Cluster of Excellence, Helmholtz-Zentrum Dresden-Rossendorf (HZDR), Dresden 01328, Germany*
[5] *School of Physical Science and Technology, Ningbo University, Ningbo 315211, China*

[†]These authors contributed equally to this work
[*]Corresponding authors, emails: dmitrii.semenok@hpstar.ac.cn (D. Semenok)
huangxiaoli@jlu.edu.cn (X. Huang)

## Abstract

High pressure plays a crucial role in the field of superconductivity. Compressed hydride superconductors are leaders in the race for a material that can conduct electricity without resistance at high or even room temperature. In the present work, we have discovered new lanthanum superhydride, cubic A15-type La$_4$H$_{23}$, with lower stabilization pressure compared to the reported *fcc*-LaH$_{10}$. Superconducting La$_4$H$_{23}$ was obtained by laser heating of LaH$_3$ with ammonia borane at about 120 GPa. Transport measurements reveal the maximum critical temperature $T_c$ (onset) = 105 K and the critical field $\mu_0 H_{C2}(0) \approx 32$ T at 118 GPa, as evidenced by the sharp drop of electrical resistance and the displacement of superconducting transitions in magnetic fields. Moreover, we provide evidence for unconventional transport associated with a pseudogap phase in La$_4$H$_{23}$ using pulsed magnetic fields up to 68 T. A large negative magnetoresistance in the non-superconducting state below 40 K, quasi *T*-linear electrical resistance, and a sign-change of its temperature dependence mark the emergence of pseudogap in this hydride. Discovered lanthanum hydride is a new member of the A15 family of superconductors with $T_C$ exceeding the boiling point of liquid nitrogen.

## Introduction

The search for high-temperature superconductors has been an important goal pursued tirelessly by researchers since the discovery of the superconductivity in mercury [1]. However, until 2014, the critical temperature ($T_c$) of conventional superconductors had never exceeded the McMillan limit (~ 40 K) [2]. As one of the most effective methods of changing the structure of matter, pressure can lead to appearance of unusual properties of materials that are unlikely to occur in ambient conditions, such as conventional high-temperature superconductivity (HTSC). The increase of $T_c$ in Hg-containing cuprates to 164 K under high pressure motivated extensive further research in this field [3]. On the basis of the chemical pre-compression idea, first proposed by Ashcroft [4], the breakthrough discovery of conventional HTSC in compressed sulfur hydride $H_3S$ with $T_c$ above 200 K at 150 GPa was made both theoretically and experimentally in 2014-2015 [5, 6]. Soon after, the HTSC record was broken by lanthanum superhydride $LaH_{10}$ with $T_c$ = 250 K at 170 GPa [7, 8]. The successful use of high pressure in the search for new superconductors provided a clear path to even room-temperature superconductivity and attracted considerable scientific interest. Over the past five years, such remarkable superconductors as $YH_6$ [9], $YH_9$ [10], $CeH_{10}$ [11] and $CaH_6$ [12] have been found.

To date, $LaH_{10}$ has the highest $T_c$ among binary hydrides, and the La-H system is one of the best studied both experimentally and theoretically [13-21]. Recently, a series of novel lanthanum hydrides was found using single-crystal X-ray diffraction analysis including traces of $Pm\bar{3}n$-$La_4H_{23}$ [18] which belongs to the so-called A15 class of superconductors. However, due to the complex composition of the mixture and the low content of this phase, the transport properties of $La_4H_{23}$ have not been studied. A15 (β-W) structural type is a series of intermetallic compounds with outstanding low-temperature superconducting properties and $Pm\bar{3}n$ space group. Examples are vanadium silicide $V_3Si$ ($T_c$ = 17 K), $Nb_3Ge$ ($T_c$ = 23 K) and $Nb_3Sn$ ($T_c$ = 18.3 K).

In this work, we successfully synthesized a new representative of this class, cubic $La_4H_{23}$ with A15 structure which stable in the pressure range of 91–120 GPa, using laser heating of $LaH_{3-x}$ with ammonia borane ($NH_3BH_3$) above 1500 K. Transport measurements showed a sharp drop of the sample electrical resistance in $10^3$ times with the highest onset $T_c$ = 105 K, and the decrease of the critical temperature under external magnetic fields. Theoretical calculations support conclusions of the experiment and point to a strong electron-phonon interaction in the hydrogen sublattice of $Pm\bar{3}n$-$La_4H_{23}$. In addition, we performed a comprehensive study of the magnetic phase diagram and magnetoresistance (MR) of A15 $La_4H_{23}$ in pulsed magnetic field up to 68 T. Our observations reveal the unconventional nature of the non-superconducting state in this lanthanum hydride.

**Experimental details**

In this work, we used powder $LaH_{3-x}$ (x ≈ 0.1–0.2) with metallic conductivity, which was preliminarily synthesized from La powder and hydrogen under pressure, and ammonia borane (AB, $NH_3BH_3$) as the hydrogen source and pressure transmitting medium, considering the high-temperature decomposition reaction: $NH_3BH_3 \rightarrow 3H_2$ + *c*-BN [11]. All the crystal structures were determined by in situ powder high-pressure X-ray diffraction (XRD) at the SPring-8 (Japan) synchrotron radiation source, beamline BL10XU, using the wavelength of 0.4124 Å. The $CeO_2$ was used as a calibrant. Dioptas software was used to analyze the experimental XRD patterns [22].

Diamond anvil cells (DACs) are essential for high-pressure experiments, and we used the DACs made of NiCrAl (40HNU) alloy. Diamond anvils had a culet of 60 μm, beveled at 8° to a diameter of 250 μm. For transport measurements, the MgO/epoxy was selected as insulting layer placed between the tungsten gasket and platinum foil leads. Molybdenum electrodes were sputtered on the surface of diamond anvils to connect the sample and Pt leads. The pressure was determined by the diamond Raman shift [23]. Further experimental details described in the Supporting Information.

Magnetoresistance measurements in pulsed magnetic fields were carried out in a 24 mm bore of 72 T resistive pulse magnet (pulse time of 150 ms) at the Helmholtz-Zentrum Dresden-Rossendorf (HLD HZDR). Strands of the Litz wire glued to the silver paint were moved closer together to minimize open loop pickup.

All twisted pairs were fixed using the GE7031 varnish. A bath helium cryostat was used, which made it possible excellent control of the DAC temperature between 4.5 K and 78 K. A 100 cm long NiCr wire with a resistance of about 150 Ohms wrapped around the diamond chamber was used as a heater. Cernox thermometers were attached to the DAC's body (60 mm long and 12 mm in diameter) for measurements of the temperature. A high-frequency (33.33 and 16.666 kHz) lock-in amplifier technique was employed to measure the sample resistance. For the measurements in high magnetic fields, we used a four-probe AC method with the excitation current of 1-2 mA (10 mA already significantly suppresses superconductivity in $La_4H_{23}$). The voltage drop across the sample was amplified by an instrumentation amplifier and detected by a lock-in amplifier. In general, we used the same methodology as in the previous studies of $(La, Nd)H_{10}$ [24], $SnH_4$ [25], and $CeH_{9-10}$ [26].

## Computational details

The evolutionary algorithm USPEX [27-29] was used to predict thermodynamically stable La-H phases. To investigate the La–H system, we performed both fixed- and variable-composition searches at 100, 120, 150 and 200 GPa. The number of generations was 100. We calculated the convex hulls in the temperature range from 0 K to 2000 K, using free energies computed by Phonopy [30]. Metastable structures with the energy ≤ 30 meV/atom above the hull are also presented on the convex hulls.

Structure relaxations and energy calculations were performed using the VASP code[31-33] within density functional theory (DFT)[34, 35], implementing the Perdew–Burke–Ernzerhof (PBE) exchange–correlation functional[36] and the projector-augmented wave (PAW) method[37, 38]. The kinetic energy cutoff was set at 600 eV. Γ-centered k-point meshes with a resolution of $2\pi \times 0.05$ Å$^{-1}$ were used for sampling the Brillouin zone. The phonon band structure and density of states were computed using Phonopy [30] package implementing the finite displacement method, $2 \times 2 \times 2$ supercells were generated. The energy cutoff and k-spacing parameters for the VASP calculations were set at 500 eV and $2\pi \times 0.1$ Å$^{-1}$, respectively. Sumo package [39] was used to visualize the phonon density of states and band structure. The k-points for phonon band structures were chosen using Hinuma's recommendation [40]. The Phonopy package was also used to calculate zero-point energy (ZPE) corrections and thermal properties, such as entropy and free energy. To calculate phonon frequencies and electron–phonon coupling (EPC) coefficients, we used Quantum Espresso (QE) package [41] utilizing density functional perturbation theory (DFPT) [42], plane-wave PZ HGH pseudopotentials and the tetrahedron method [43, 44]. In general, we used the same methodology as in the study of La-Mg-H system [45].

## Results and discussion

*Superconducting properties of $La_4H_{23}$*

The scheme of the electrical DAC H1 prepared for the transport measurements is shown in Figures 1b, c. Laser heating of the $LaH_3$/AB sample to temperature above 1500 K was performed at 123 GPa, then the pressure reduced to 120 GPa. Cryogenic measurements of the sample immediately showed the appearance of a drop in electrical resistance from 0.12 Ω to $10^{-4}$ Ω at 93 K (onset) corresponding to the manifestation of superconductivity in the sample (Figure 1a). After the second laser heating, a sharp superconducting transition was observed at 90 K at 114 GPa. The third laser heating led to an increase in pressure to 118 GPa and the critical temperature $T_c$ also rose to 105 K. This value is very close to the critical temperatures of cerium superhydrides $CeH_9$ and $CeH_{10}$ in the same pressure range [11], but the advantage of the studied $La_4H_{23}$ is the much smaller amount of hydrogen (< 6 atoms per La) required to obtain this result. To further investigate the dependence of $T_c$ on pressure, we decompressed the DAC to 91 GPa (Figure 1a). We found that $T_c$ $(P)$ decreases monotonically during the decompression runs reaching the maximum of 105 K at 118 GPa (Figure 1c).

Remarkably, a change in the sign of the quasi-linear temperature dependence of the electrical resistance ($dR/dT$) is observed during decompression (Figure 1a) of the DAC H1. This phenomenon was previously observed during decompression in sulfur hydrides $H_2S$ and $H_3S$ [6], in phosphorus hydride $PH_3$ [46], in $CeH_{10}$ [26] and in ternary superhydride $(La,Ce)H_9$ [47]. As we have shown earlier for $SnH_4$ [48], the change in the sign of $dR/dT$ is incompatible with the view of $La_4H_{23}$ as a normal Fermi-liquid metal whose resistance is due to the scattering of electrons on phonons. Indeed, the influence of electron-phonon interaction on the transport properties of metals can be described using the Eliashberg transport spectral function $\alpha^2 F_{tr}(\omega)$, which, as a rule, differs little from the Eliashberg function for electron-phonon interaction $\alpha^2 F(\omega)$ [49]. In the first approximation of the variational solution of the Boltzmann equations for electron transport [50], the dependence of electrical resistivity ($\rho$) on temperature ($T$) is linear

$$\rho(T) = T \frac{\pi V_{cell} k_B}{N_F \langle v_x^2 \rangle} \int_0^\infty \frac{d\omega}{\omega} \frac{x^2}{sh^2(x)} \alpha^2 F_{tr}(\omega), \qquad (1)$$

where $x = \hbar\omega/k_B T$, $V_{cell}$ – is the unit cell volume, $N_F$ – is the electron density of states at the Fermi level and the $\langle v_x^2 \rangle$ – is the band-averaged Fermi speed of electrons. All of the above parameters are positive in all known three-dimensional materials, which leads to positive $dR/dT$ for the vast majority of metals and alloys. Very rare exceptions are complex alloys with disorder effects at low temperatures, such as manganin and constantan[51-53]. At high temperatures the term $\frac{x^2}{sh^2(x)} \to 1$, and we come to simple formula (2)

$$\rho(T) = \lambda_{tr} \frac{\pi V_{cell} k_B}{N_F \langle v_x^2 \rangle} T, \qquad (2)$$

where $\lambda_{tr}$ – is the transport electron-phonon coupling (EPC) parameter [49]. From this equation it is absolutely clear that if $dR/dT < 0$, then the transport EPC parameter must also be negative ($\lambda_{tr} < 0$), and the usual electron-phonon coupling strength ($\lambda$) will also be negative or about zero, which is incompatible with the concept of conventional superconductivity in $La_4H_{23}$. The negative sign of $dR/dT$ and the narrowing of superconducting transitions in a magnetic field were independently confirmed by Cross et al.[54] and, therefore, are reproducible phenomena.

Instead, the properties of polyhydrides in the non-superconducting state on the verge of their dynamic stability [55] should be described in the framework of a non-Fermi liquid model close to the models developed for describing the pseudogap phase and metal-to-insulator transitions in cuprates [56]. As we will see in the next paragraph, A15 $La_4H_{23}$ exhibits the strong negative magnetoresistance characteristic of superconductors with a pseudogap phase. As can be seen from Figure 1c, a further decrease in pressure of DAC H1 leads to broadening of superconducting transitions up to complete disappearance of the superconducting state and decomposition of $La_4H_{23}$ (Figures 1c, d).

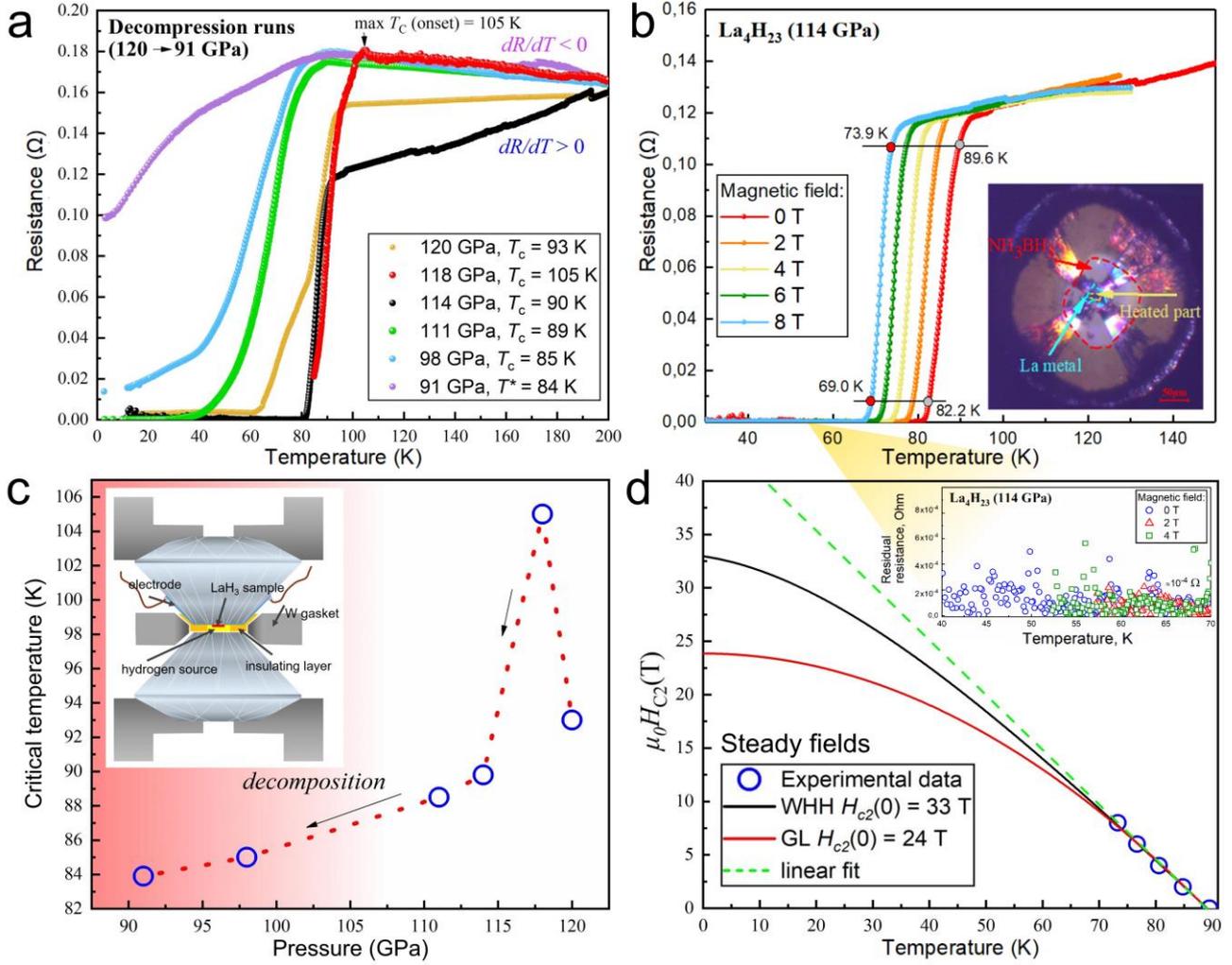

**Figure 1.** Electrical DAC construction and transport properties of $La_4H_{23}$ under pressure. (a) Dependence of the electrical resistance of the sample on temperature during decompression of DAC H1 from 120 GPa to 91 GPa. $T_c$ corresponds to the onset transition point. (b) Dependence of electrical resistance of the sample on applied magnetic field at 114 GPa. Inset: photograph of the sample loaded in the DAC's chamber and four Mo electrodes after the laser heating at 114 GPa. The hydrogen source ($NH_3BH_3$) and the $LaH_3$ are indicated by red and blue curves, respectively. (c) Pressure dependence of the critical temperature of $La_4H_{23}$. Inset: schematic diagram of the electrical DAC with the four-electrode van der Pauw scheme. (d) Upper critical magnetic field $B_{C2}(T)$ of $La_4H_{23}$ at 114 GPa obtained by extrapolation of the experimental data (steady fields) using the Ginzburg-Landau (GL), [57] the Werthamer-Helfand-Hohenberg (WHH) [58, 59] and linear models. Inset: residual resistance of the sample after SC transition.

To further confirm the superconductivity, we measured the electrical resistance of the sample in DAC H1 in steady magnetic fields ranging from 0 to 8 T. Figure 1e shows the temperature dependence of electrical resistance of A15 $La_4H_{23}$ in steady magnetic fields at 114 GPa. The sample demonstrates the absence (and even negative value) of broadening of superconducting transitions in a magnetic field, as previously observed for yttrium ($YH_6$) [60] and lanthanum-yttrium ($(La,Y)H_{10}$) hydrides [61]. $T_c$ shifts linearly to lower temperatures as the magnetic field increases from 0 to 8 T, as it should for superconductors. To estimate the upper critical magnetic field $\mu_0H_{C2}(0)$, we applied the GL model [57], and the WHH model [58], simplified by Baumgartner [59]. As Figure 1f shows, these two models yield the $\mu_0H_{C2}(0)$ as 24 T and 33 T, respectively.

The coherence length can be estimated by $\mu_0 H_{C2} = \Phi_0/(2\pi\xi^2)$ [62]. The $\xi_{WHH}(0)$ and $\xi_{GL}(0)$ are equal to 3.16 nm and 3.72 nm, respectively. These values are higher than the same parameters in *fcc* LaH$_{10}$ [62].

Interestingly, the superconducting properties of La$_4$H$_{23}$ are found to be more pronounced than those of the recently studied A15 Lu$_4$H$_{23}$ (max $T_c$ = 71 K, [63]). This indicates that isostructural lutetium compounds, such as proposed LuH$_{10}$, LuH$_9$, and LuH$_6$, will have lower critical temperatures than the corresponding lanthanum polyhydrides in contradiction with earlier theoretical predictions [64]. Moreover, due to the smaller atomic radius of Lu, we would have to apply much higher pressures to stabilize the corresponding lutetium hydrides (e.g., LuH$_{10}$), making them less convenient to study than LaH$_x$.

*Pulsed magnetic field experiments and crystal structure of La$_4$H$_{23}$*

Unusual transport properties of La$_4$H$_{23}$ were the reason for further studies of this compound in DAC H2 at 121 GPa in stronger pulsed magnetic fields up to 68 T (Figure 2). Before the experiment, the DAC H2 was tested in steady fields up to 8T (Supporting Figure S8). The sample exhibited a superconducting transition at 84 K (onset), the extrapolated $\mu_0 H_{C2}(0)$ is 32-45 T. Thus, the reproducibility of the synthesis results was demonstrated, although in this particular sample the derivative *dR/dT* was positive.

Pulse measurements allowed us to establish the presence of a region of pronounced negative magnetoresistance at temperatures below 40 K and at the field *H* > 25 T, and obtain inflection points where MR changes sign back to positive (Figure 2a, b). These points limit a separate region of the magnetic phase diagram (MR < 0, Figure 2b), which corresponds to the pseudogap phase of A15 La$_4$H$_{23}$. The amplitude of the jump in electrical resistance (Re Z) $R_{max}/R_n$ = 1.55 at 4.5 K in the pseudogap region exceeds that in cerium superhydride CeH$_{10}$, for which we previously found similar phenomena [26]. As in the case of cerium polyhydrides, a sign reversal of $dR_{max}/dT$ is observed in the pseudogap region: $R_{max}$ reaches maximum of 0.185 Ω at 4.5 K and decreases along with temperature to $R_n$ = 0.115 Ω at 50.9 K. It is interesting to note that above the superconducting transition, at 102 K, the phenomenon of zero magnetoresistance is observed: MR ∝ *dR(H)/dH* = 0 (Supporting Figure S10).

We also were able to completely suppress superconductivity at $\mu_0 H_{C2}(0) \approx$ 32 T in the La$_4$H$_{23}$ and complete the magnetic phase diagram for this compound (Figure 2b). In general, the behavior of $\mu_0 H_{C2}(T)$, is in satisfactory agreement with the WHH model traditionally used for Bardeen-Cooper-Schrieffer superconductors.

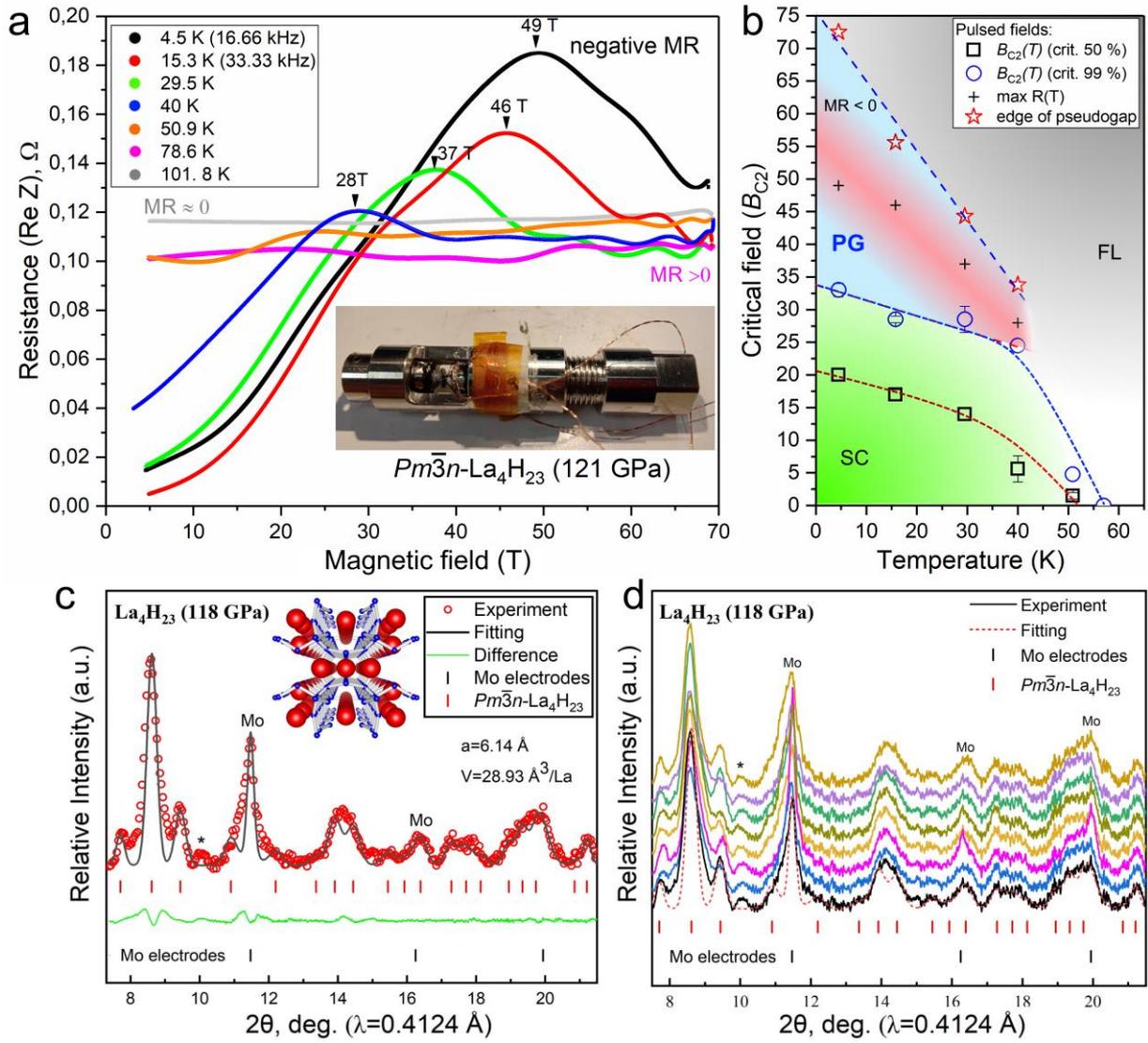

**Figure 2.** Transport properties in pulsed magnetic field and X-ray diffraction analysis of $La_4H_{23}$. (a) Dependence of the electrical resistance of the sample in DAC H2 on the external magnetic field $R(H)$ measured in the AC mode at frequencies of 16.66 kHz (4.5 K) and 33.33 kHz (most cases). For ease of analysis, the original data were smoothed using a Fourier filter (Origin lab). Raw data can be found in the Supporting Information. Inset: photo of the DAC H2. (b) Magnetic phase diagram of $La_4H_{23}$ at 121 GPa. "SC" marks superconducting region, the pseudogap phase ("PG") is marked by red and blue color where MR < 0, "FL" denotes the Fermi liquid metal behavior of the sample with MR > 0 and positive $dR/dT > 0$. (c) X-ray diffraction pattern and Le Bail refinement of the unit cell parameters of $La_4H_{23}$ phase at 118 GPa. Experimental XRD data and the Le Bail fit are represented by red hollow circles and black lines, respectively. XRD pattern contains spurious signals from molybdenum (Mo) electrodes. (d) XRD patterns measured across the sample with a step of 5 μm. The sample is very homogeneous.

For the first time among polyhydrides, structural type A15 ($Pm\bar{3}n$) was found when studying the formation of europium hydrides [65] above 100 GPa. Then, polyhydrides with the same structure were found in Ba-H system [66], among lanthanum [18] and lutetium [63] polyhydrides and in the Y-H system [67], forming a fairly large family of superhydrides. As we found out in the previous section, A15 $La_4H_{23}$ has the highest $T_c$ among all superconductors with A15 structure. Below we discuss the results of X-ray diffraction analysis of this compound.

Considering the large number of superconducting phases in the La-H system at megabar pressures, we combined the experimental powder XRD data and the computational crystal structure search [18]. The in-situ synchrotron X-ray diffraction patterns, shown in Figure 2c, indicate that the lanthanum sublattice of DAC H1 sample possessing cubic $Pm\bar{3}n$ symmetry at 118 GPa. The volume of the unit cell is 28.9 Å$^3$/La and the unit cell parameter a = 6.14 Å (Z = 8). This is in close agreement with the theoretical results in 118 GPa (≈ 28 Å$^3$/La, see Supporting Figure S12). We also compared obtained unit cell volume of $Pm\bar{3}n$-La$_4$H$_{23}$ with the reported one [18], which is 27.98 Å$^3$/La at 150 GPa. This is slightly different due to higher pressure in the Laniel et al. experiment. The hydrogen content can be estimated by the difference between atomic volumes of La atoms and the hydrogen [68, 69]. According to Ref. [70] the volume of a hydrogen atom in pure compressed hydrogen at 118 GPa is 2.20 Å$^3$/H, whereas La atom has a volume of 15.7 Å$^3$/La[68]. Combining these data, we conclude that the La:H composition in our compound is close to 1:6.

Figure 2d shows a series of XRD patterns measured at several positions across the sample in steps of 5 μm. All XRD images are almost identical at all locations containing the signals from the Mo electrodes and $Pm\bar{3}n$-La$_4$H$_{23}$. This confirms the uniform distribution of hydrogen and selective formation of only one phase in the process of synthesis. Computer modeling of this structure shows that in the hydrogen cage of La$_4$H$_{23}$, the shortest H-H distance is about 1.3 Å at 118 GPa, which is in the range of 1.0 – 1.5 Å, typical for hydride superconductors. This bond length is much longer than in pure H$_2$: $d_{HH}$ ≈ 1.1 Å at 115 GPa [22].

*Theoretical analysis*

To confirm the experimental observations and find the most probable hydrogen sublattice in the synthesized lanthanum hydride, we performed a series of thermodynamic calculations of the enthalpies of formation of various La polyhydrides at 100, 120, 150, and 200 GPa and temperatures from 0 to 2000 K, considering the zero-point energy (ZPE) contribution in the harmonic approximation. The corresponding convex hulls are shown in Supporting Figures S1-S6. As can be seen, $Pm\bar{3}n$-La$_4$H$_{23}$ is close to the stability region and about 20-30 meV/atom above the convex hull at 100-150 GPa. This phase is dynamically stable at 120 GPa already in the harmonic approximation (Figure 3b), although further pressure reduction to 100 GPa destabilizes it in agreement with the experiment (Figure 1a). Increasing the temperature has a little effect on its relative stability (Supporting Figures S4-S6), whereas the stability of the neighboring highly symmetric phases $Im\bar{3}m$-LaH$_6$ and $Fm\bar{3}m$-LaH$_{10}$ is strongly affected. We found that $Im\bar{3}m$-LaH$_6$ is dynamically unstable in the harmonic approximation at 100-120 GPa and can undergo a distortion $Im\bar{3}m$ → $Cmme$ → $C2/m$ at low pressure, which stabilizes La$_4$H$_{23}$. This is also supported by the known experimental fact of distortion of $Fm\bar{3}m$-LaH$_{10}$ at pressures below 140 GPa [17]. Thus, the convex hull deformation at low pressures, caused by a much faster destabilization of the $Im\bar{3}m$-LaH$_6$, may lead to the possibility of experimental observation of $Pm\bar{3}n$-La$_4$H$_{23}$. Indeed, all lanthanum hydrides discussed above were synthesized in recent experiments [18].

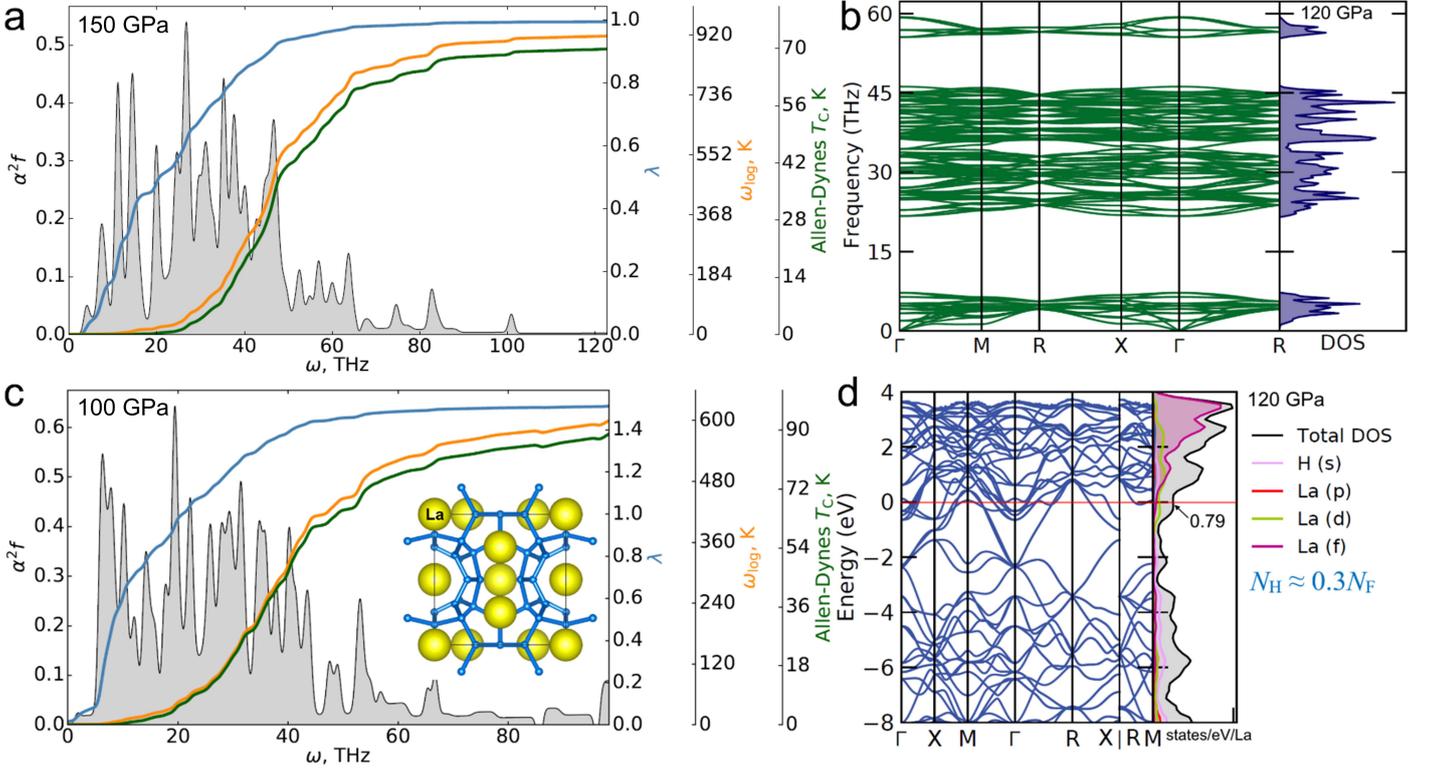

**Figure 3.** Results of theoretical calculations of electronic, phonon and superconducting properties of $La_4H_{23}$ at 100, 120 and 150 GPa. (a, c) Eliashberg functions of $La_4H_{23}$ calculated without considering symmetry at 150 and 100 GPa, respectively. We used a k-mesh of 8×8×8 and a q-mesh of 2×2×2. The results were also verified on k-mesh 12×12×12 and q-mesh 3×3×3. Pictures were prepared using python script available on GitHub [71]. (b) Phonon band structure and density of states of $La_4H_{23}$ at 120 GPa calculated in the harmonic approximation. The compound was found to be dynamically stable at this pressure. (d) Electron band structure and density of states projected on H and La atoms at 120 GPa. The contribution of hydrogen is ≈ 30 % of the total density of states.

Superconducting properties of $Pm\bar{3}n$-$La_4H_{23}$ have been investigated theoretically using the norm-conserving Goedecker-Hartwigsen-Hutter (HGH) pseudopotential with the Perdew-Zunger (LDA) functional and *k, q*-grids of low density. These pseudopotentials give results that are in close agreement with the experimental data, so it is this computational approach that we will discuss below (Table 1 and Figure 3 a, c). We found that $La_4H_{23}$ exhibits moderate superconducting properties at 100 GPa, and the electron-phonon interaction strength reaches λ ≈ 1.5. Critical temperature of superconductivity decreases with increasing pressure due to a decrease in the λ. The calculated $T_c$ of $La_4H_{23}$ reaches 92-95 K at 100 GPa in the harmonic approximation in agreement with the experiment. The relatively low $T_c$ of this superhydride correlates with the low contribution of the hydrogen sublattice to the density of electronic states (Figure 3d), which is equal to about 30% of the total density of electronic states at the Fermi level ($N_F$ = 0.79 states/eV/La). The expected [72] upper critical magnetic field $μ_0H_{C2}(0)$ = 29-51 T is also in reasonable agreement with the experiment (Figure 2b).

**Table 1.** Parameters of the superconducting state of $Pm\bar{3}n$-La$_4$H$_{23}$ calculated at 100 and 150 GPa in the harmonic approximation using HGH PZ pseudopotentials. "McM" stands for the McMillan formula [2], "A-D" stands for the Allen-Dynes model [73], and "E" corresponds to the solution of the isotropic Eliashberg equations [74]  ($\mu^* = 0.1$ in all cases).

| Pressure | 100 GPa | 150 GPa |
|---|---|---|
| $\lambda$ | 1.49 | 1.0 |
| $\omega_{log}$, K | 650 | 892 |
| $\omega_2$, K | 1121 | 1383 |
| $T_c$ (McM) | 73.5 | 61.4 |
| $T_c$ (A-D) | 91.7 | 68.0 |
| $T_c$ (E) | 94.7 | 70.4 |
| $N_F$, states/spin/Ry/Å$^3$ | 0.196 | 0.182 |
| $2\Delta/k_BT_c$ | 4.69 | 4.04 |
| Expected $\mu_0H_{C2}(0)$, T | 51 | 29 |
| $2\Delta$, meV | 19.1 | 12.2 |

Thus, as can be clearly seen, theoretical calculations show excellent agreement with experimental data for La$_4$H$_{23}$. However, this situation should be considered as unique! To demonstrate it, we carried out a primary analysis of the superconducting properties of two other polyhydrides of the A15 family: Lu$_4$H$_{23}$ and Y$_4$H$_{23}$. Calculations at 200 GPa for A15 Lu$_4$H$_{23}$ give $T_c$ (A-D) = 187 K, $\lambda$ = 2.32 and $\omega_{log}$ = 867 K (Supporting Figure S14). This $T_c$ is almost 3 times greater than the experimental result $T_c$ (exp) = 70 K[63], which we independently verified and confirmed[72].

To date, there are several publications that are in qualitative agreement with the results of our calculations for Lu$_4$H$_{23}$. First of all, it's prediction of the room-temperature superconductivity in LuH$_6$ [64] at 100 GPa. Considering the similar hydrogen content in Lu$_4$H$_{23}$ and LuH$_6$, as well as the high symmetry of these hydrides, it is not difficult to guess that in the experiment LuH$_6$ will have a much lower $T_c$ than predicted by theory. Secondly, one cannot fail to mention predictions of the room-temperature superconductivity in the Lu-Y-H system at a pressure of 100-200 GPa [75]. In both of these cases, we are dealing with a clear deviation of theoretical predictions from the experiment, due to strong correlation effects in *f*-shell of Lu, the inclusion of which has an extremely strong effect on superconductivity, but cannot be carried out within the framework of DFT.

A very similar situation is observed for $Pm\bar{3}n$-Y$_4$H$_{23}$ at 150 GPa: calculations with different *q*-point meshes lead to the fact that $T_c$(Y$_4$H$_{23}$) must exceed 250 K (Supporting Figures S15-S16). This is hard to believe, given that previously studied $Im\bar{3}m$-YH$_6$ also demonstrated an abnormally low critical temperature: being a predicted room-temperature superconductor with $T_c$ > 270 K [76], it exhibits superconductivity only below 224 K [9]. Thus, here we encounter the problem of discrepancy between DFT calculations and experimental results that currently has no solution and should be addressed to future research.

## Conclusions

We have synthesized novel lanthanum superhydride $La_4H_{23}$ with the A15 structure via laser heating of $LaH_3$ and $NH_3BH_3$ at 118-123 GPa. Powder X-ray diffraction revealed a $Pm\bar{3}n$-$La_4H_{23}$ phase as the main reaction product, which has a clathrate hydrogen sublattice. Transport measurements confirm pronounced superconducting properties of $La_4H_{23}$ with a maximum transition temperature of 105 K at 118 GPa in agreement with theoretical calculations. The calculated electron-phonon interaction strength λ is around 1.5. During decompression, $T_c$ of $La_4H_{23}$ decreases along with the pressure below 118 GPa, and at 91 GPa the superconducting transition completely disappears. In the non-superconducting state between 98 and 120 GPa, $La_4H_{23}$ demonstrates a change in the sign of the quasi-linear temperature dependence of the electrical resistance corresponding to the non-Fermi liquid (strange metal) behavior.

Experimental magnetic phase diagrams down to 4.5 K was constructed and the upper critical field of $La_4H_{23}$ $\mu_0H_{C2}(0)$ = 32 T were established. The experimental behavior of $\mu_0H_{C2}(T)$ in this compound, in general, can be described in terms of the WHH model. Discovered $La_4H_{23}$ exhibits pronounced properties of a strange metal below $T_c$: large *H*-linear negative magnetoresistance, the reversal of the sign of temperature coefficient of electrical resistance (*dR/dT*) below 40 K, and quasi-linear *R(T)* in non-superconducting state, which corresponds to the properties of the pseudogap phase of cuprate superconductors. The experiment shows that the physics of high-$T_C$ superhydrides is rather close to this of cuprates.

## Author statement

**Jianning Guo**: Validation, formal analysis, investigation, visualization, writing of original draft. **Grigoriy Shutov**: Validation, theoretical analysis, investigation, visualization. **Di Zhou**: Investigation. **Su Chen**: Investigation. **Yulong Wang**: Investigation. **Toni Helm**: Investigation. **Sven Luther**: Investigation. **Dmitrii Semenok**: Investigation, writing of original draft, theoretical calculations, review & editing. **Xiaoli Huang**: Investigation, writing, review & editing, funding acquisition. **Tian Cui**: Writing and review & editing, funding acquisition.

## Declaration of competing interest

The authors declare no competing financial interest or personal relationships that could have appeared to influence the work reported in this paper.

## Data availability

The authors declare that the main data supporting the findings of this study are contained within the paper and its associated Supporting Information. All other relevant data are available from the corresponding author upon request.

## Acknowledgements

Authors thank the staff of SPring-8 Synchrotron Radiation Facility for their help during the synchrotron XRD measurements. The authors express their gratitude to Prof. Viktor Struzhkin (HPSTAR) for financial support.

**Funding**

This work was supported by National Key R&D Program of China (2022YFA1405500), the National Natural Science Foundation of China (11974133 and 52072188), the Program for Changjiang Scholars and Innovative Research Team in University (IRT_15R23), and users with Excellence Program of Hefei Science Center CAS 2021HSC-UE011, Zhejiang Provincial Science and technology innovation Team (2021R01004), and Jilin Provincial Science and Technology Development Project (20210509038RQ), the Fundamental Research Funds for the Central Universities. D. S. and D. Z. thank National Natural Science Foundation of China (NSFC, grant No. 1231101238) and Beijing Natural Science Foundation (grant No. IS23017) for support of this research. This work was supported by HLD HZDR, member of the European Magnetic Field Laboratory (EMFL).

# SUPPORTING INFORMATION

# Large negative magnetoresistance and pseudogap phase in superconducting A15-type La$_4$H$_{23}$

Jianning Guo [1,†], Dmitrii Semenok [2,†,*], Grigoriy Shutov [3,†], Di Zhou[2], Su Chen[1], Yulong Wang[1], Toni Helm[4], Sven Luther[4], Xiaoli Huang[1,*], and Tian Cui [1,5]

[1] *State Key Laboratory of Superhard Materials, College of Physics, Jilin University, Changchun 130012, China*

[2] *Center for High Pressure Science and Technology Advanced Research (HPSTAR), Beijing*

[3] *Skolkovo Institute of Science and Technology, Skolkovo Innovation Center, 3 Nobel Street, Moscow 143026, Russia*

[4] *Hochfeld-Magnetlabor Dresden (HLD-EMFL) and Würzburg-Dresden Cluster of Excellence, Helmholtz-Zentrum Dresden-Rossendorf (HZDR), Dresden 01328, Germany*

[5] *School of Physical Science and Technology, Ningbo University, Ningbo 315211, China*

[†]These authors contributed equally to this work
[*]Corresponding authors, emails: <dmitrii.semenok@hpstar.ac.cn> (D. Semenok)
<huangxiaoli@jlu.edu.cn> (X. Huang)

# Content

# 1. Thermodynamic calculations

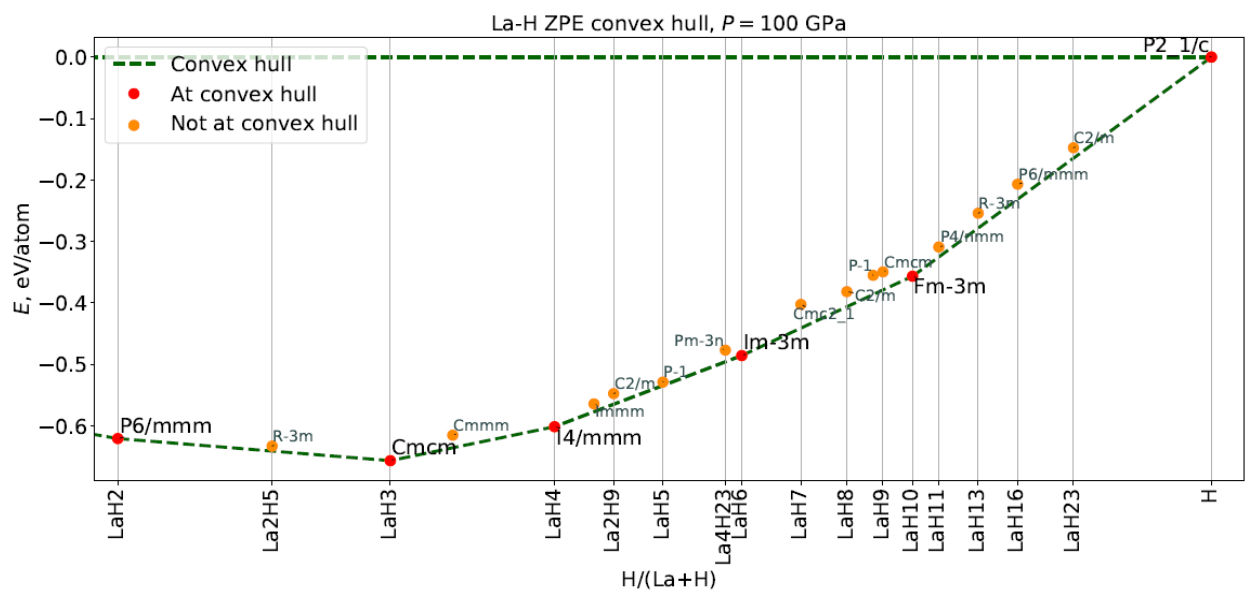

**Figure S1.** Convex hull of the La-H system calculated at 100 GPa and 0 K in harmonic approximation. Only those phases that lie less than 30 meV/atom from the convex hull are shown. Zero-point energy (ZPE) was included in the calculations.

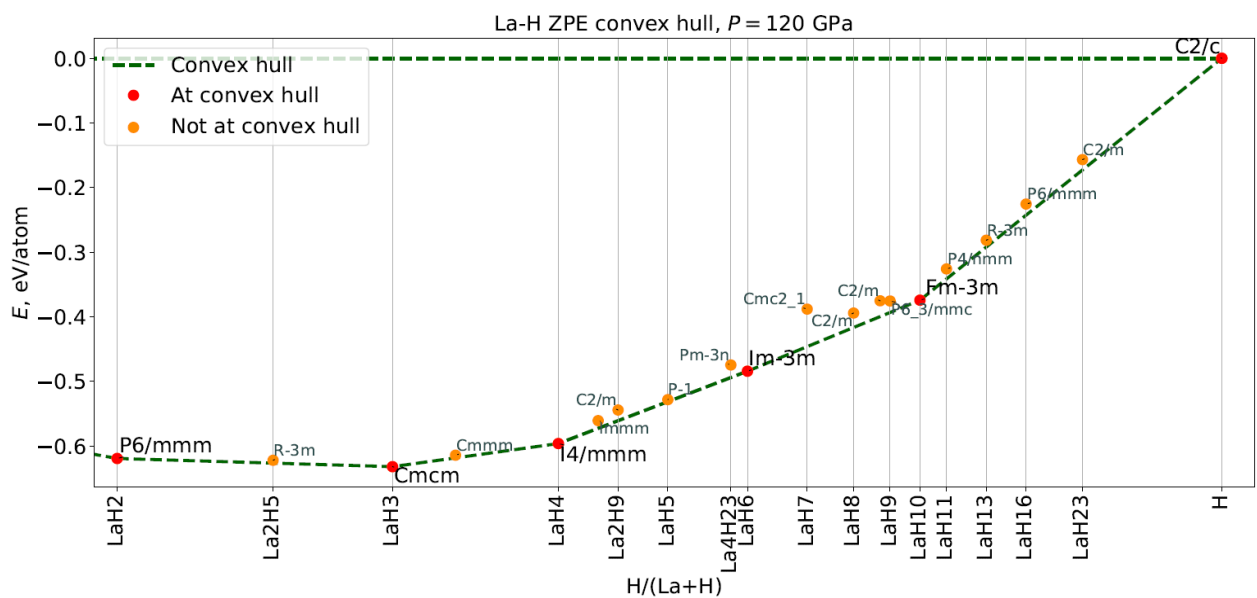

**Figure S2.** Convex hull of the La-H system calculated at 120 GPa and 0 K in harmonic approximation. Only those phases that lie less than 30 meV/atom from the convex hull are shown. Zero-point energy (ZPE) was included in the calculations.

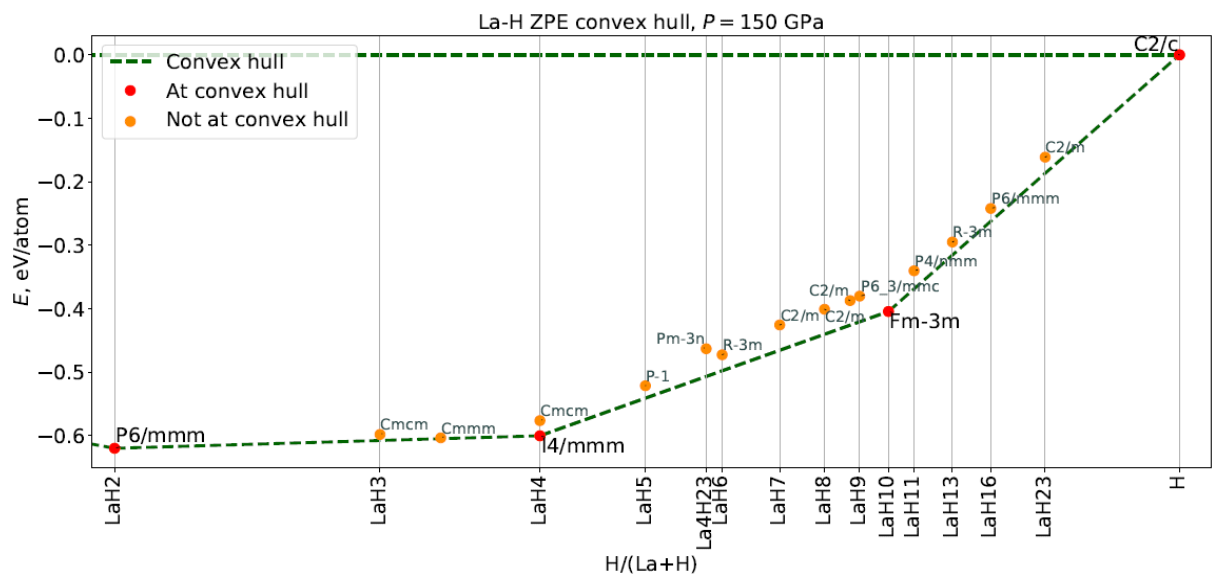

**Figure S3.** Convex hull of the La-H system calculated at 150 GPa and 0 K in harmonic approximation. Only those phases that lie less than 30 meV/atom from the convex hull are shown. Zero-point energy (ZPE) was included in the calculations.

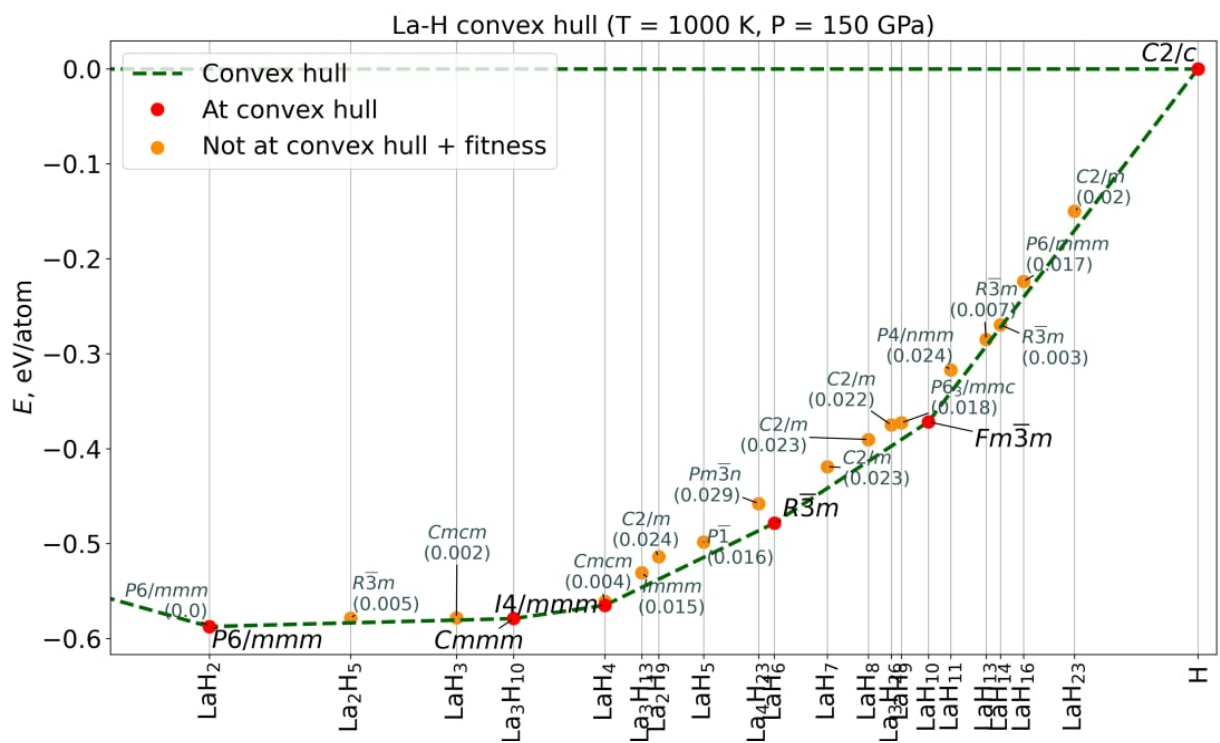

**Figure S4.** Convex hull of the La-H system calculated at 150 GPa and 1000 K in harmonic approximation. Only those phases that lie less than 30 meV/atom from the convex hull are shown. Zero-point energy (ZPE) was included in the calculations. The numbers in parentheses correspond to the distance of the connections from the convex hull in eV/atom.

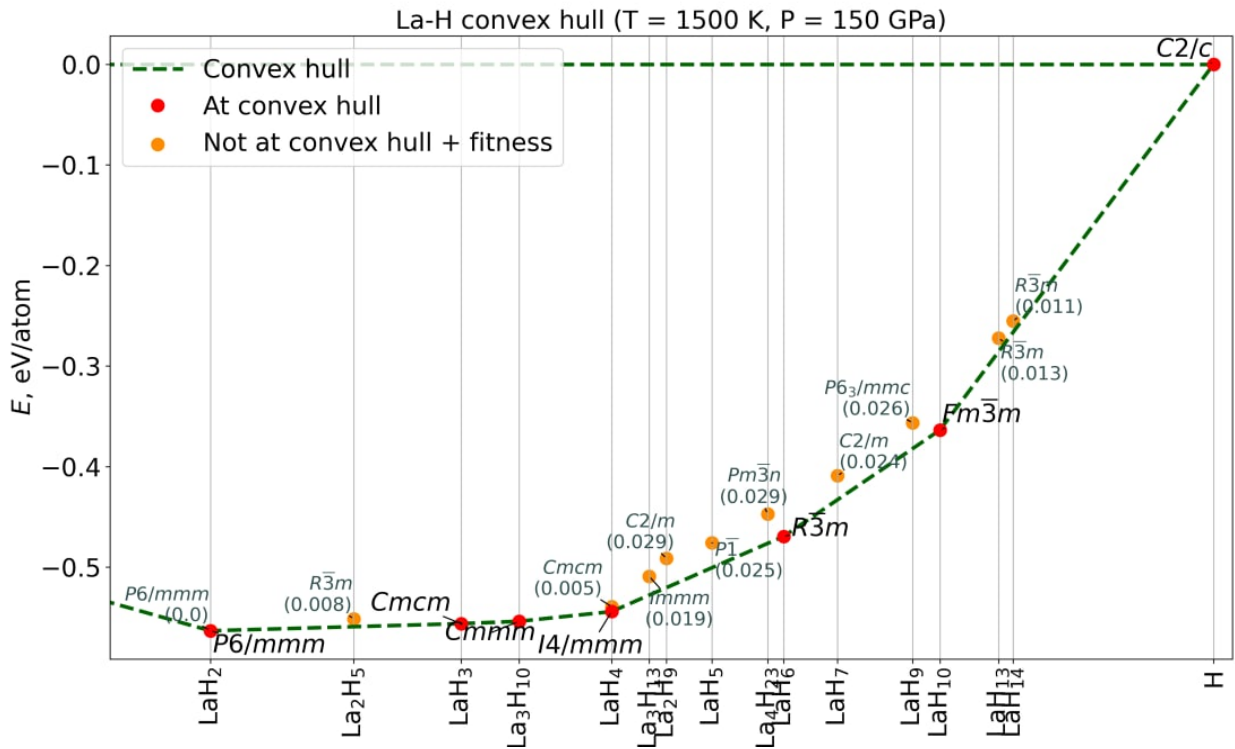

**Figure S5.** Convex hull of the La-H system calculated at 150 GPa and 1500 K in harmonic approximation. Only those phases that lie less than 30 meV/atom from the convex hull are shown. Zero-point energy (ZPE) was included in the calculations. The numbers in parentheses correspond to the distance of the connections from the convex hull in eV/atom.

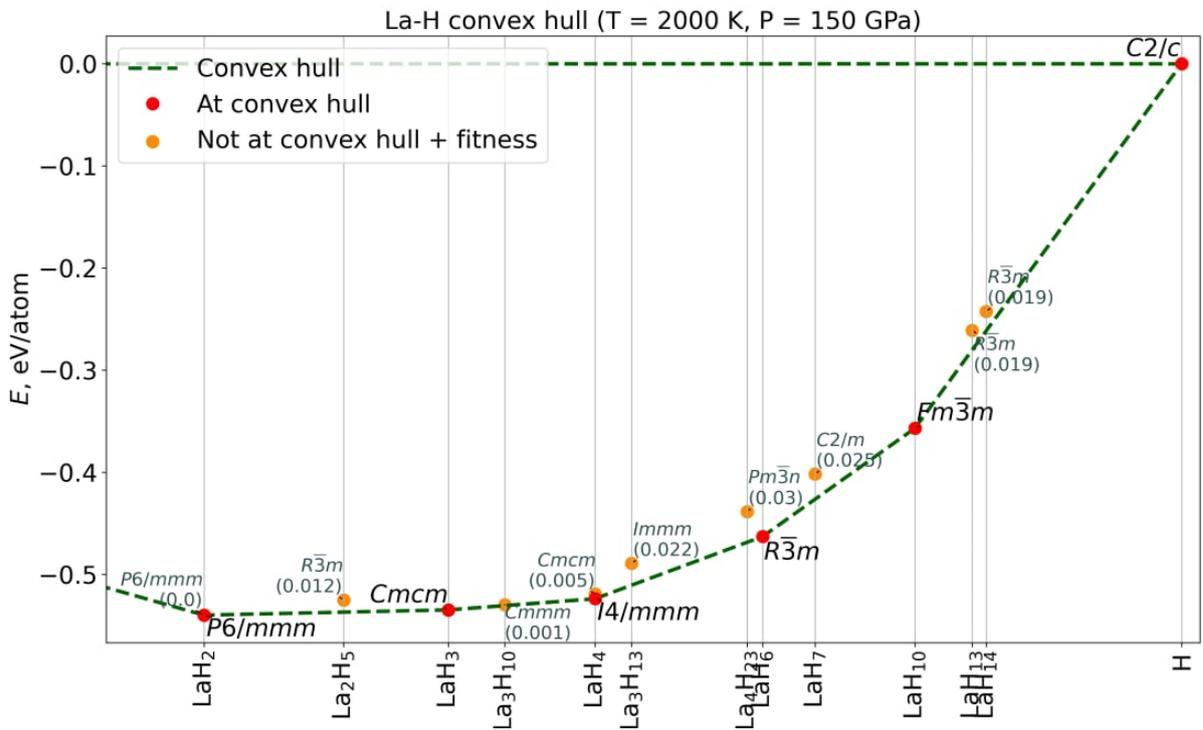

**Figure S6.** Convex hull of the La-H system calculated at 150 GPa and 2000 K in harmonic approximation. Only those phases that lie less than 30 meV/atom from the convex hull are shown. Zero-point energy (ZPE) was included in the calculations. The numbers in parentheses correspond to the distance of the connections from the convex hull in eV/atom.

**Table S1.** Enthalpies of formation of various lanthanum hydrides (in eV/atom) at 100 GPa and 0 K (Figure S1). ZPE stands for the zero-point energy. X, Y are coordinates of lanthanum hydrides on the two-dimensional convex hull diagram. Fitness is the distance from the convex hull line.

| Space Group | Formula | X | E | ZPE | E+ZPE | Y | ZPE Y | Fitness without ZPE | ZPE fitness |
|---|---|---|---|---|---|---|---|---|---|
| Fm-3m | LaH$_{10}$ | 0,90900 | -0,62980 | 0,15270 | -0,47710 | -0,26900 | -0,35680 | 0,03340 | 0,00 |
| C2/m | La$_2$H$_9$ | 0,81800 | -0,04020 | 0,19930 | 0,15910 | -0,52800 | -0,54790 | 0,01110 | 0,01730 |
| P4/nmm | LaH$_{11}$ | 0,91700 | -0,72460 | 0,22610 | -0,49850 | -0,29300 | -0,30920 | 0,00 | 0,01620 |
| C2/m | LaH$_{23}$ | 0,95800 | -0,96460 | 0,24850 | -0,71610 | -0,14400 | -0,14760 | 0,00280 | 0,01710 |
| R-3m | LaH$_{13}$ | 0,92900 | -0,77100 | 0,21920 | -0,55180 | -0,22830 | -0,25420 | 0,02910 | 0,02420 |
| C2/m | LaH$_8$ | 0,88900 | -0,52480 | 0,20630 | -0,31840 | -0,35250 | -0,38190 | 0,01520 | 0,02460 |
| P-1 | La$_3$H$_{26}$ | 0,89700 | -0,57270 | 0,21120 | -0,36150 | -0,32890 | -0,35530 | 0,01810 | 0,03130 |
| Cmcm | LaH$_9$ | 0,90 | -0,60650 | 0,21910 | -0,38730 | -0,33050 | -0,34970 | 0,01110 | 0,02950 |
| P6/mmm | LaH$_{16}$ | 0,94100 | -0,83340 | 0,21400 | -0,61940 | -0,17310 | -0,20700 | 0,03810 | 0,02430 |
| Pm-3n | La$_4$H$_{23}$ | 0,85200 | -0,26240 | 0,18580 | -0,07660 | -0,43590 | -0,47710 | 0,02510 | 0,01910 |
| Cmcm | LaH$_3$ | 0,75 | 0,47480 | 0,19560 | 0,67040 | -0,64960 | -0,65710 | 0,00 | 0,00 |
| R-3m | La$_2$H$_5$ | 0,71400 | 0,83010 | 0,18910 | 1,01920 | -0,62760 | -0,63330 | 0,00210 | 0,00820 |
| I4/mmm | LaH$_4$ | 0,80 | 0,08260 | 0,18800 | 0,27060 | -0,57490 | -0,60180 | 0,00 | 0,00 |
| P6/mmm | LaH$_2$ | 0,66700 | 1,29180 | 0,17300 | 1,46480 | -0,61050 | -0,62110 | 0,00 | 0,00 |
| Fm-3m | La | 0,00 | 8,12600 | 0,02700 | 8,15290 | 0,00 | 0,00 | 0,00 | 0,00 |
| Cmc2_1 | LaH$_7$ | 0,87500 | -0,44290 | 0,23010 | -0,21290 | -0,40030 | -0,40270 | 0,20 | 0,03860 |
| Immm | La$_3$H$_{13}$ | 0,81200 | -0,00250 | 0,19660 | 0,19410 | -0,54340 | -0,56460 | 0,00670 | 0,01280 |
| P2_1/c | H | 1,00 | -1,20950 | 0,26180 | -0,94770 | 0,00 | 0,00 | 0,00 | 0,00 |
| Im-3m | LaH$_6$ | 0,85700 | -0,30910 | 0,17550 | -0,13360 | -0,43320 | -0,48600 | 0,01630 | 0,00 |
| Cmmm | La$_3$H$_{10}$ | 0,76900 | 0,34050 | 0,19630 | 0,53690 | -0,60430 | -0,61560 | 0,01310 | 0,02050 |
| P-1 | LaH$_5$ | 0,83300 | -0,15900 | 0,19900 | 0,04 | -0,50540 | -0,52910 | 0,00 | 0,00570 |

**Table S2.** Enthalpies of formation of various lanthanum hydrides (in eV/atom) at 120 GPa and 0 K (Figure S2). ZPE stands for the zero-point energy. X, Y are coordinates of lanthanum hydrides on the two-dimensional convex hull diagram. Fitness is the distance from the convex hull line.

| Space Group | Formula | X | E | ZPE | E+ZPE | Y | ZPE Y | Fitness without ZPE | ZPE fitness |
|---|---|---|---|---|---|---|---|---|---|
| Fm-3m | LaH$_{10}$ | 0,90900 | -0,22280 | 0,16470 | -0,05800 | -0,29430 | -0,37440 | 0,02550 | 0,00 |
| C2/m | La$_2$H$_9$ | 0,81800 | 0,54580 | 0,20900 | 0,75480 | -0,53020 | -0,54440 | 0,01120 | 0,01650 |
| C2/m | La$_3$H$_{26}$ | 0,89700 | -0,13350 | 0,21010 | 0,07660 | -0,34360 | -0,37530 | 0,02230 | 0,02450 |
| P4/nmm | LaH$_{11}$ | 0,91700 | -0,32380 | 0,23210 | -0,09160 | -0,31160 | -0,32600 | 0,00 | 0,01550 |
| C2/m | LaH$_{23}$ | 0,95800 | -0,62820 | 0,25530 | -0,37290 | -0,15580 | -0,15680 | 0,80 | 0,01600 |
| P6_3/mmc | LaH$_9$ | 0,90 | -0,16390 | 0,20280 | 0,03900 | -0,33590 | -0,37560 | 0,01610 | 0,01780 |
| C2/m | LaH$_8$ | 0,88900 | -0,07070 | 0,21110 | 0,14050 | -0,36550 | -0,39430 | 0,01600 | 0,02240 |
| Immm | La$_3$H$_{13}$ | 0,81200 | 0,59440 | 0,20540 | 0,79980 | -0,54440 | -0,56080 | 0,00730 | 0,01190 |
| R-3m | LaH$_{13}$ | 0,92900 | -0,39150 | 0,21570 | -0,17580 | -0,24790 | -0,28150 | 0,02240 | 0,01060 |
| Pm-3n | La$_4$H$_{23}$ | 0,85200 | 0,26420 | 0,19630 | 0,46050 | -0,43980 | -0,47470 | 0,02830 | 0,01940 |
| Cmcm | LaH$_3$ | 0,75 | 1,20050 | 0,20370 | 1,40430 | -0,62870 | -0,63210 | 0,00 | 0,00 |
| R-3m | La$_2$H$_5$ | 0,71400 | 1,60400 | 0,19630 | 1,80030 | -0,61980 | -0,62220 | 0,00210 | 0,00440 |
| I4/mmm | LaH$_4$ | 0,80 | 0,70150 | 0,19800 | 0,89950 | -0,57540 | -0,59630 | 0,00 | 0,00 |
| P6/mmm | LaH$_2$ | 0,66700 | 2,13760 | 0,18030 | 2,31790 | -0,61240 | -0,61940 | 0,00 | 0,00 |
| Fm-3m | La | 0,00 | 10,11560 | 0,02970 | 10,14530 | 0,00 | 0,00 | 0,00 | 0,00 |
| P6/mmm | LaH$_{16}$ | 0,94100 | -0,47160 | 0,21550 | -0,25620 | -0,18870 | -0,22560 | 0,02920 | 0,01710 |
| Cmc2_1 | LaH$_7$ | 0,87500 | 0,06060 | 0,23610 | 0,29670 | -0,38760 | -0,38820 | 0,02420 | 0,05810 |
| Im-3m | LaH$_6$ | 0,85700 | 0,20360 | 0,19 | 0,39360 | -0,44190 | -0,48430 | 0,01430 | 0,00 |
| Cmmm | La$_3$H$_{10}$ | 0,76900 | 1,01130 | 0,20290 | 1,21420 | -0,60550 | -0,61430 | 0,00120 | 0,00420 |
| P-1 | LaH$_5$ | 0,83300 | 0,39880 | 0,20830 | 0,60710 | -0,50980 | -0,52830 | 0,00 | 0,00320 |
| C2/c | H | 1,00 | -0,93280 | 0,26620 | -0,66660 | 0,00 | 0,00 | 0,00 | 0,00 |

**Table S3.** Enthalpies of formation of various lanthanum hydrides (in eV/atom) at 150 GPa and 0 K (Figure S3). ZPE stands for the zero-point energy. X, Y are coordinates of lanthanum hydrides on the two-dimensional convex hull diagram. Fitness is the distance from the convex hull line.

| Space Group | Formula | X | E | ZPE | E+ZPE | Y | ZPE Y | Fitness without ZPE | ZPE fitness |
|---|---|---|---|---|---|---|---|---|---|
| *P-1* | LaH$_5$ | 0,83300 | 1,18930 | 0,21930 | 1,40860 | -0,50900 | -0,52140 | 0,00 | 0,01960 |
| *C2/c* | H | 1,00 | -0,55310 | 0,27 | -0,28300 | 0,00 | 0,00 | 0,00 | 0,00 |
| *Fm-3m* | LaH$_{10}$ | 0,90900 | 0,35600 | 0,16370 | 0,51970 | -0,31900 | -0,40440 | 0,01730 | 0,00 |
| *Cmcm* | LaH$_4$ | 0,80 | 1,57860 | 0,21760 | 1,79620 | -0,57 | -0,57640 | 0,00450 | 0,02390 |
| *Cmcm* | LaH$_3$ | 0,75 | 2,22530 | 0,21300 | 2,43830 | -0,59870 | -0,59820 | 0,00530 | 0,00950 |
| *I4/mmm* | LaH$_4$ | 0,80 | 1,57510 | 0,19720 | 1,77230 | -0,57350 | -0,60030 | 0,00 | 0,00 |
| *P6/mmm* | LaH$_2$ | 0,66700 | 3,33740 | 0,18570 | 3,52300 | -0,61230 | -0,62 | 0,00 | 0,00 |
| *Fm-3m* | La | 0,00 | 12,95520 | 0,04 | 12,99510 | 0,00 | 0,00 | 0,00 | 0,00 |
| *C2/m* | La$_3$H$_{26}$ | 0,89700 | 0,48300 | 0,22040 | 0,70340 | -0,36140 | -0,38720 | 0,01800 | 0,03880 |
| *R-3m* | LaH$_{13}$ | 0,92900 | 0,14100 | 0,22960 | 0,37050 | -0,27080 | -0,29490 | 0,01490 | 0,02060 |
| *P4/nmm* | LaH$_{11}$ | 0,91700 | 0,24290 | 0,24020 | 0,48320 | -0,32970 | -0,34030 | 0,00 | 0,02850 |
| *Pm-3n* | La$_4$H$_{23}$ | 0,85200 | 1,01160 | 0,20940 | 1,22100 | -0,43650 | -0,46310 | 0,03480 | 0,04370 |
| *C2/m* | LaH$_{23}$ | 0,95800 | -0,15650 | 0,26540 | 0,10900 | -0,16620 | -0,16120 | 0,40 | 0,02540 |
| *P6$_3$/mmc* | LaH$_9$ | 0,90 | 0,45140 | 0,21310 | 0,66460 | -0,34630 | -0,38020 | 0,02290 | 0,04040 |
| *C2/m* | LaH$_8$ | 0,88900 | 0,56940 | 0,22200 | 0,79140 | -0,37850 | -0,40090 | 0,01360 | 0,03940 |
| *P6/mmm* | LaH$_{16}$ | 0,94100 | 0,02690 | 0,22920 | 0,25610 | -0,21460 | -0,24200 | 0,01910 | 0,02020 |
| *C2/m* | LaH$_7$ | 0,87500 | 0,74220 | 0,20890 | 0,95110 | -0,39330 | -0,42560 | 0,02710 | 0,03990 |
| *R-3m* | LaH$_6$ | 0,85700 | 0,92860 | 0,21260 | 1,14120 | -0,44810 | -0,47260 | 0,01130 | 0,02530 |
| *Cmmm* | La$_3$H$_{10}$ | 0,76900 | 1,96300 | 0,21470 | 2,17760 | -0,60130 | -0,60350 | 0,00 | 0,00140 |

## 2. Electron, phonon, elastic and superconducting properties of La$_4$H$_{23}$

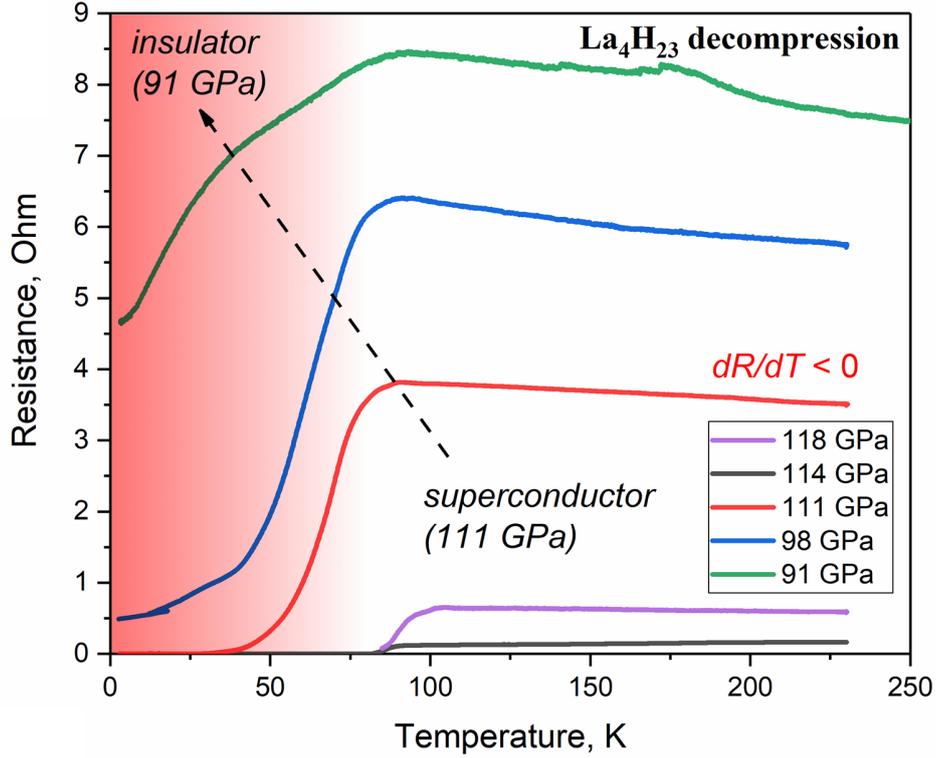

**Figure S7.** Experimental transport properties of La$_4$H$_{23}$ in DAC H1. (a) Decompression of the DAC H1 and temperature dependence of the electrical resistivity of the sample in real scale. Pronounced increase in electrical resistance in 10-100 times at $T > T_C$ is observed during decompression due to the beginning of the metal-insulator transition. The anomaly at 114 GPa is an exception. (b) Residual resistance of the sample detected in the SC state in magnetic fields of 0-4 T.

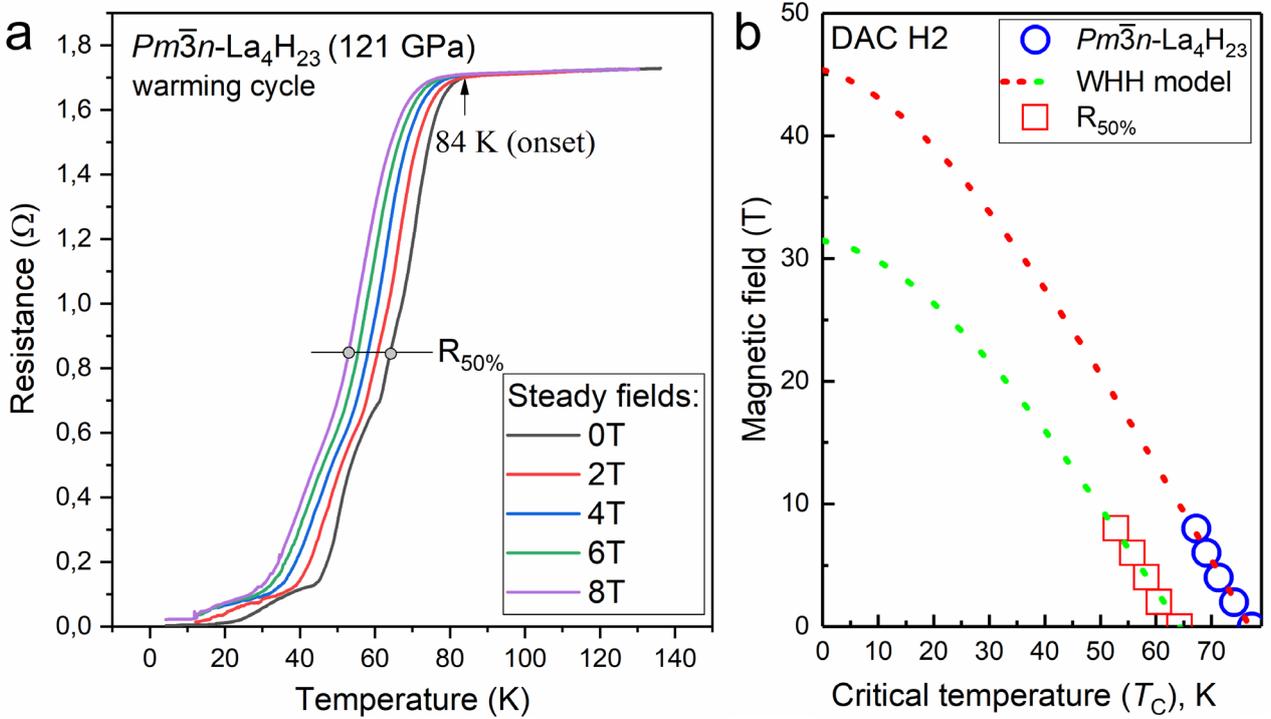

**Figure S8.** Experimental transport properties of La$_4$H$_{23}$ in DAC H2 before the pulsed field study. (a) Temperature dependence of electrical resistance of La$_4$H$_{23}$ at 121 GPa measured in magnetic fields 0-8 T in warming cycles. (b) Upper critical magnetic field $\mu_0 H_{C2}(T)$ dependence on temperature reconstructed using onset $T_C$ and $T_C$ found via R$_{50\%}$ criteria. Extrapolations based on the WHH model[1] are shown by red and green dashed lines.

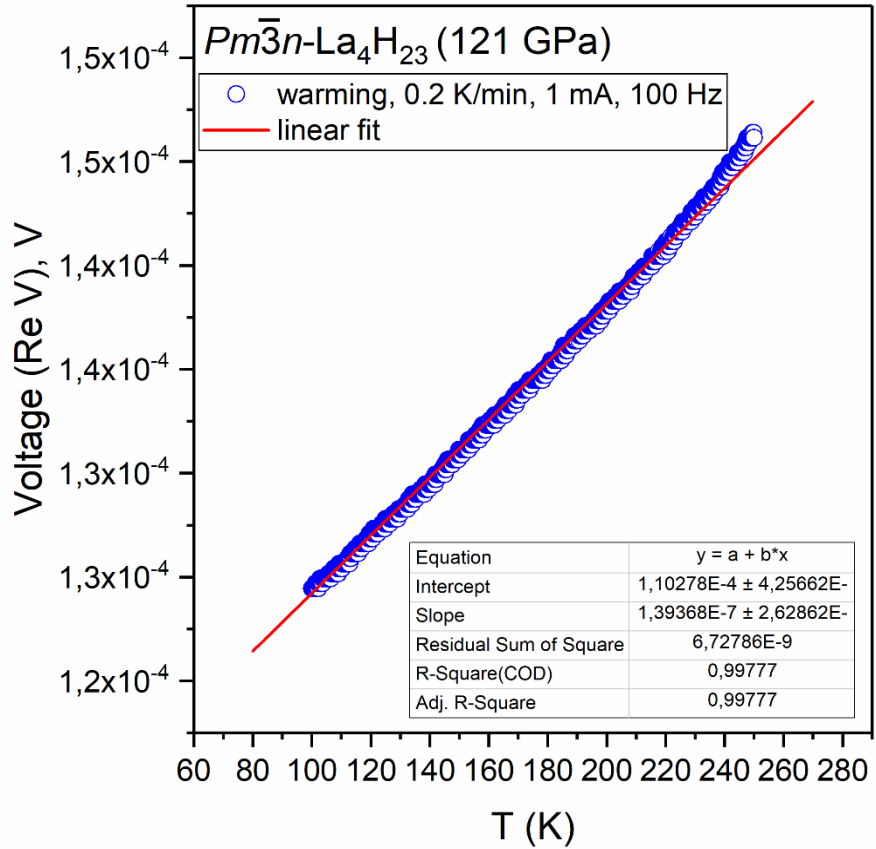

**Figure S9.** Quasi *T*-linear dependence of the voltage drop across the DAC H2 sample during 4-contact measurements in AC mode on temperature in the range from 100 K to 260 K. In fact, this *R(T)* curve cannot be explained within the framework of the electron-phonon scattering model (e.g., Bloch-Grüneisen formula) and indicates the non-Fermi liquid behavior of the sample.

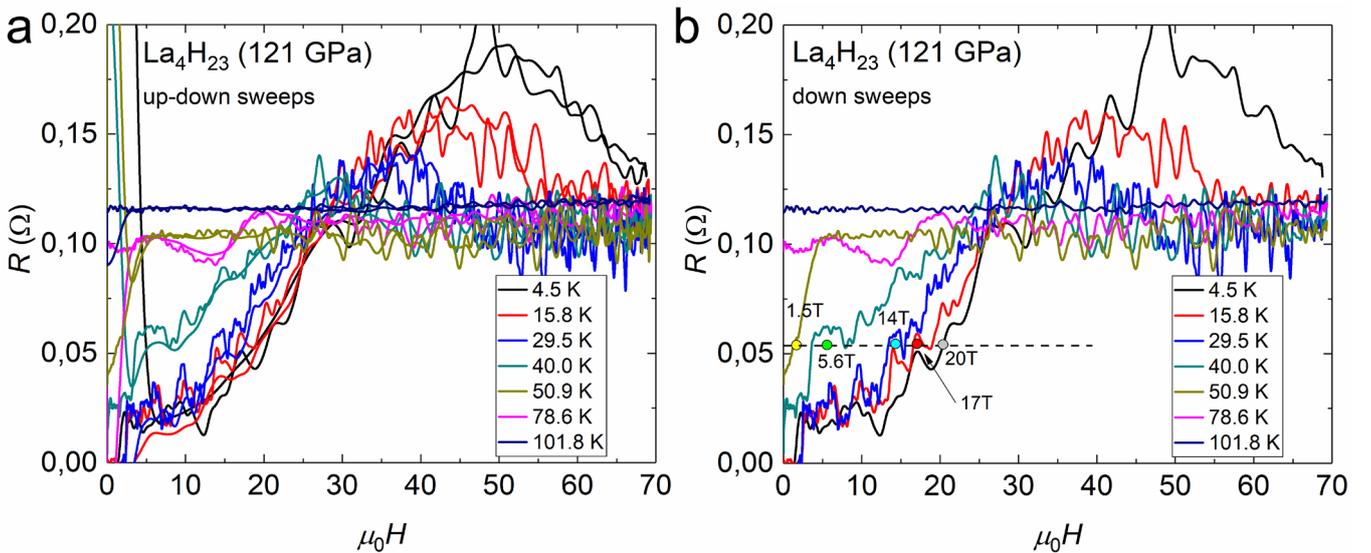

**Figure S10.** Dependence of the electrical resistance of the sample in DAC H2 on the external magnetic field *R(H)* measured in the AC mode at frequencies of 16.66 kHz (4.5 K) and 33.33 kHz (most cases). Raw data. Such loud periodic noise is caused by large open loops in the contact wires of the DAC H2. (a) Magnetic field sweeps up and down together; (b) only sweeps down are shown and used to determine $\mu_0 H_{C2}(T)$ according to $R_{50\%}$ criteria.

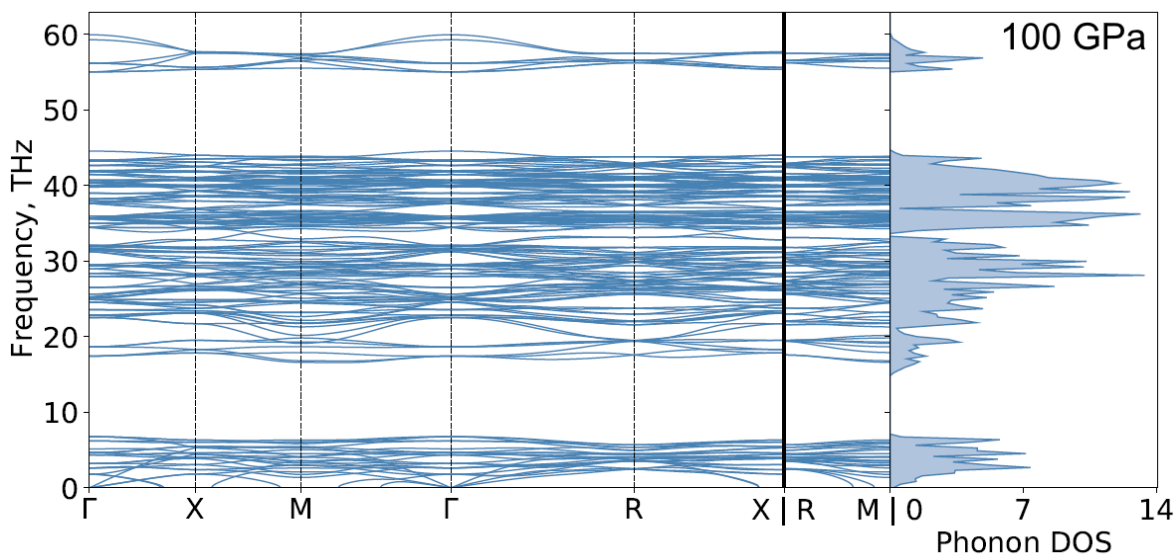

**Figure S11**. Phonon band structure and corresponding density of states of $La_4H_{23}$ at 100 GPa in harmonic approximation. The compound has a small number of imaginary phonon modes, which are likely to disappear when anharmonic effects are considered.

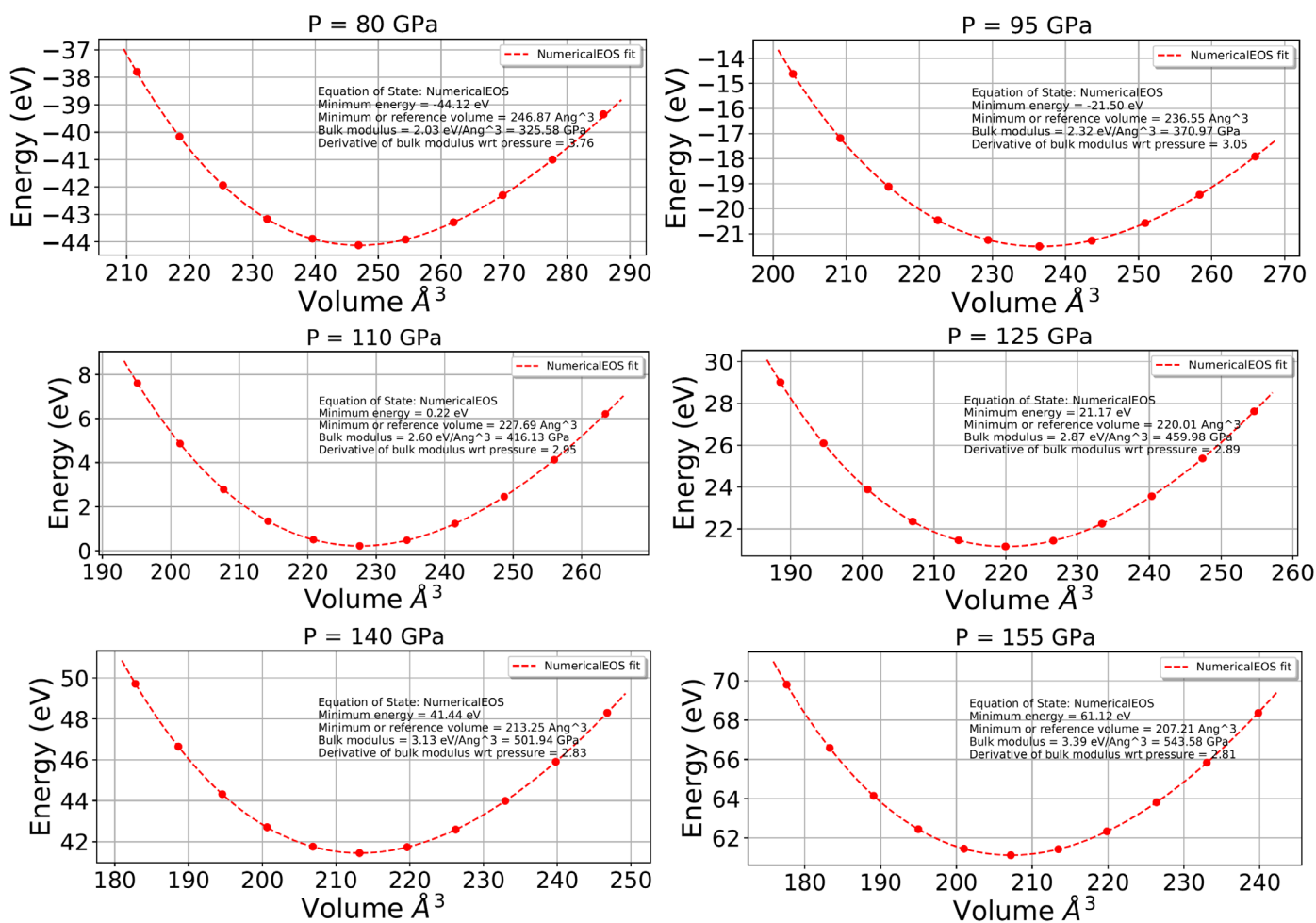

**Figure S12**. Equation of state, bulk modulus (*B*) and its derivative (*dB/dP*) of $La_4H_{23}$ at pressures from 80 to 155 GPa. Calculations were done using VASP code with the PAW PBE pseudopotentials for La and H.

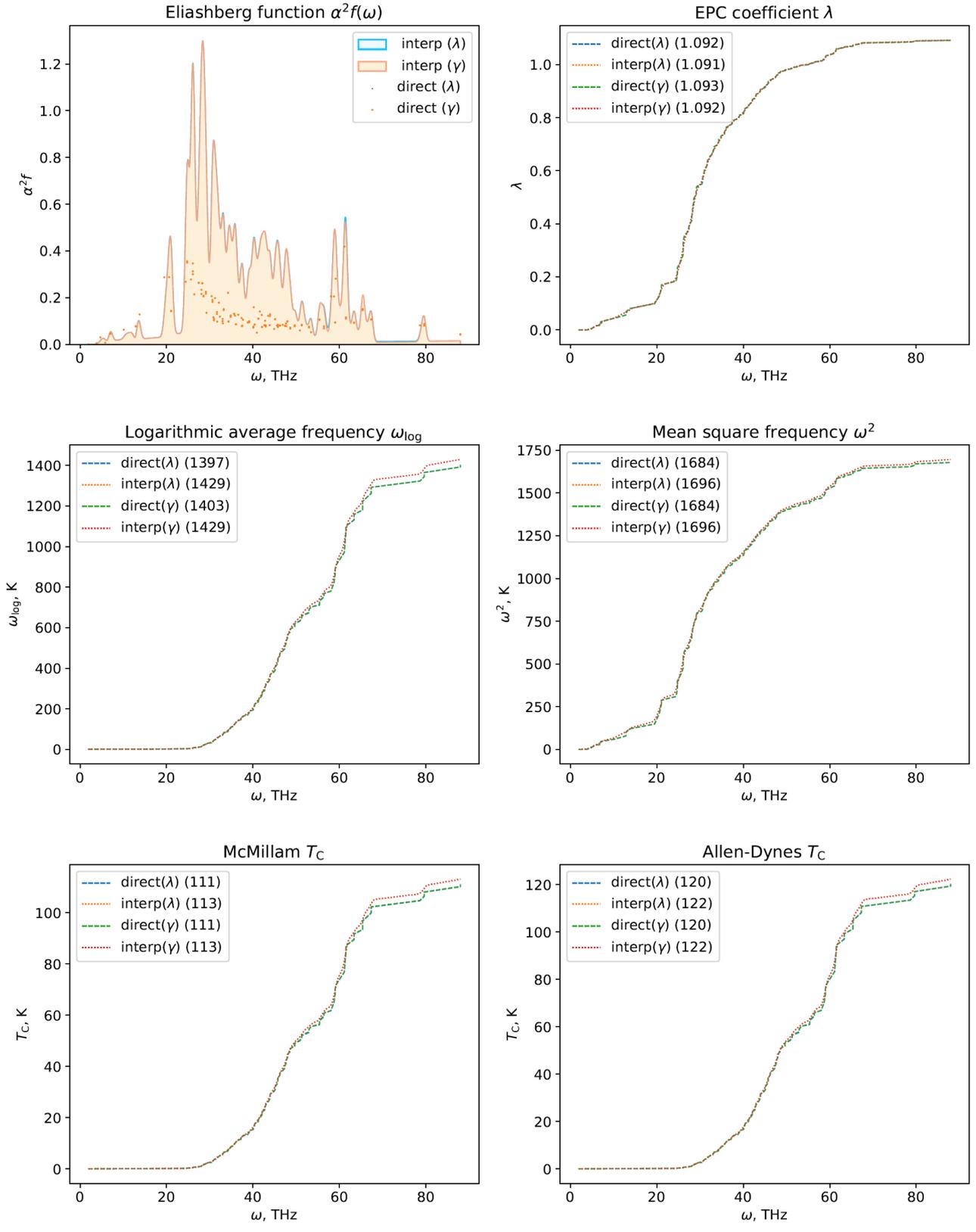

**Figure S13.** Superconducting state parameters and Eliashberg function for $La_4H_{23}$ (54 atoms in the unit cell) at 150 GPa calculated using PBE HGH pseudopotentials for La and H atoms, $q$-mesh was 2×2×2. Pictures were generated via python script available on GitHub[2].

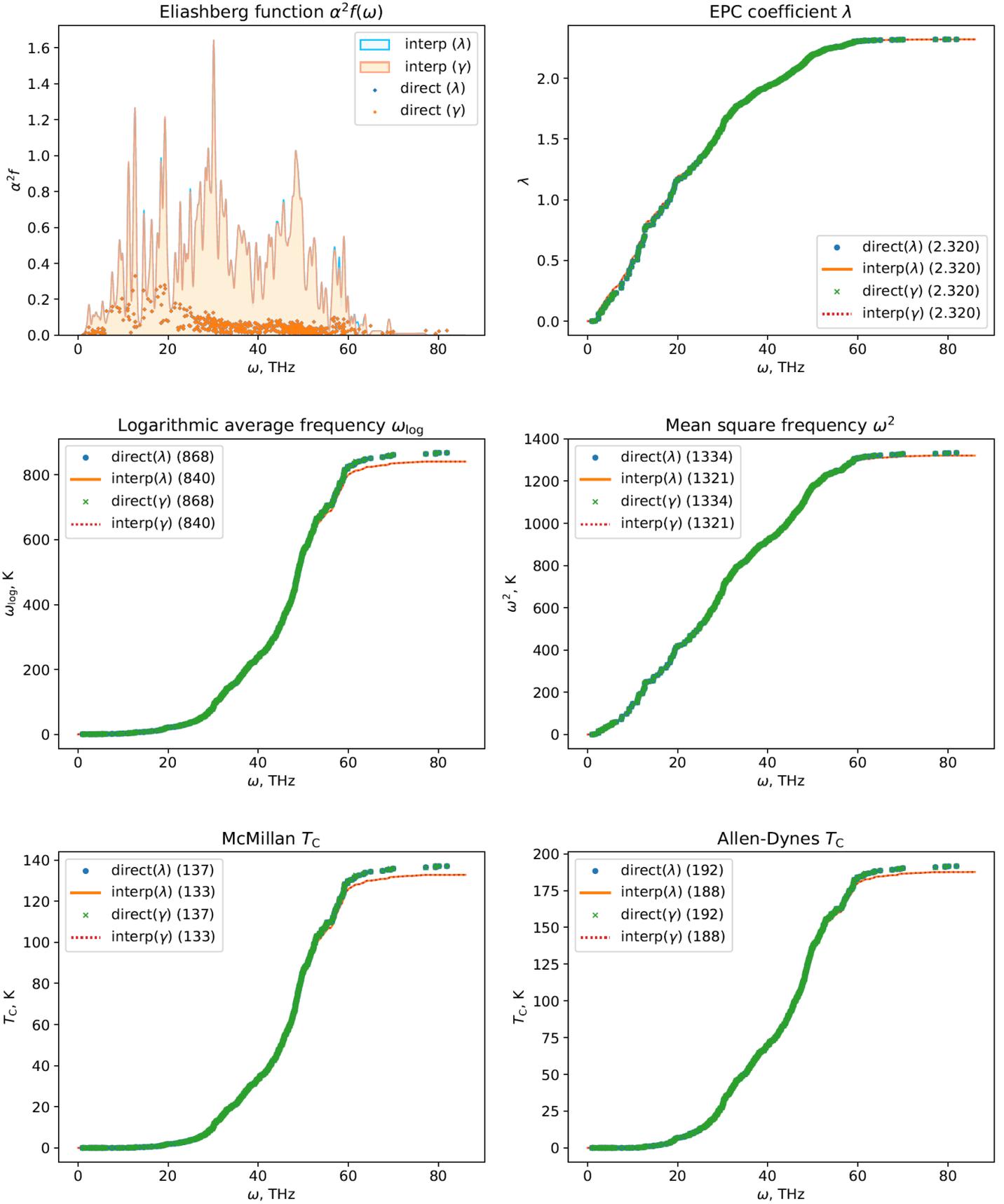

**Figure S14.** Superconducting state parameters and Eliashberg function for $Lu_4H_{23}$ (54 atoms in the unit cell) at 200 GPa calculated using PBE pseudopotentials for Lu (Lu.pbe-spdn-kjpaw_psl.1.0.0.UPF) and H atoms, $k$-mesh was 12×12×12, $q$-mesh was 4×4×4. Pictures were generated via python script available on GitHub[2].

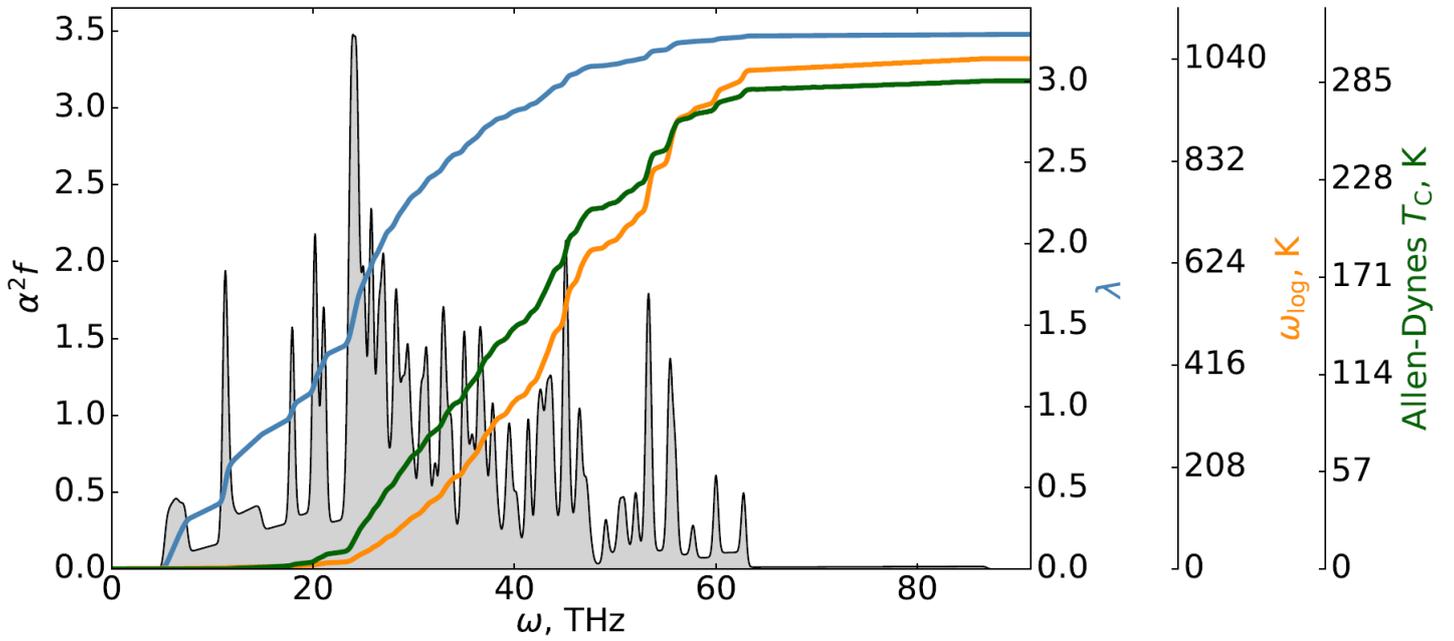

**Figure S15.** Superconducting state parameters and Eliashberg function for $Y_4H_{23}$ (54 atoms in the unit cell) at 150 GPa calculated using PBE pseudopotentials for Y (Y.pbe-spn-kjpaw_psl.1.0.0.UPF) and H atoms, *k*-mesh was 8×8×8, *q*-mesh was 2×2×2. Pictures were generated via python script available on GitHub[2].

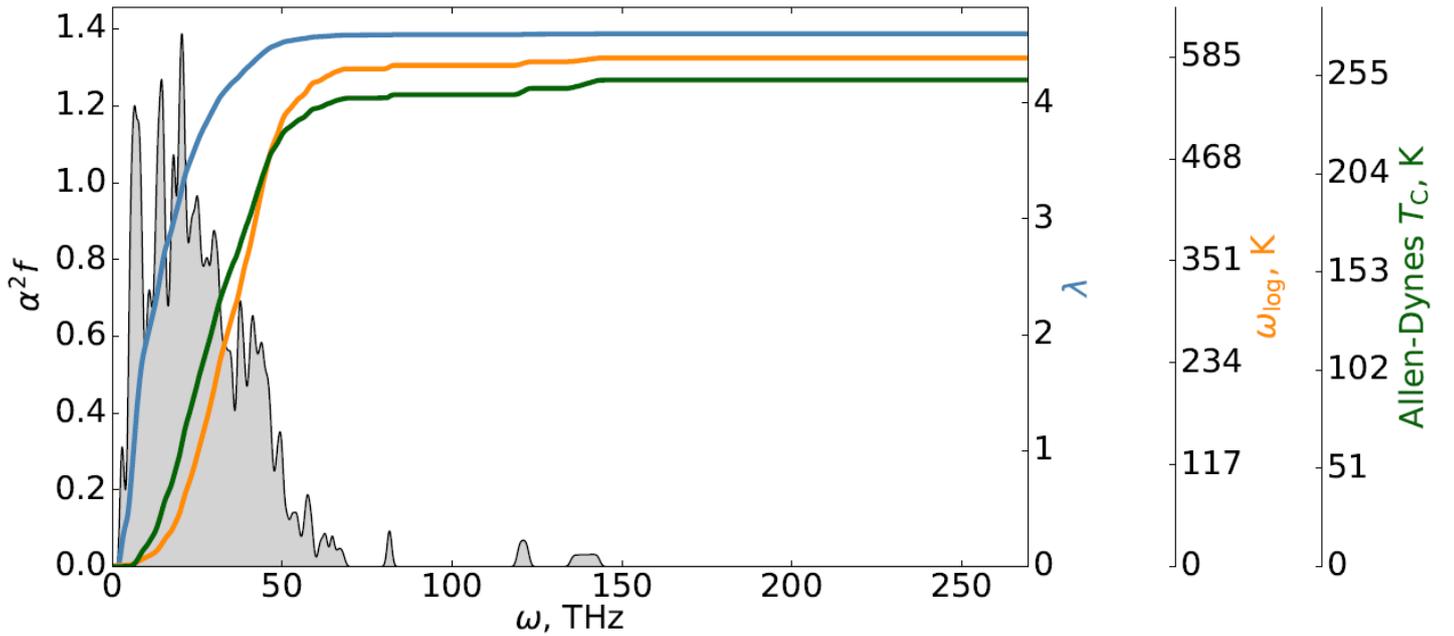

**Figure S16.** Superconducting state parameters and Eliashberg function for $Y_4H_{23}$ (54 atoms in the unit cell) at 150 GPa calculated using PBE pseudopotentials for Y (Y.pbe-spn-kjpaw_psl.1.0.0.UPF) and H atoms, *k*-mesh was 8×8×8, *q*-mesh was 4×4×4. Pictures were generated via python script available on GitHub[2].